\documentclass[twocolumn,aps,prl]{revtex4}

\usepackage{graphicx}

\begin{document}
\title{Bell's theorem without inequalities and without unspeakable information}
\author{Ad\'{a}n Cabello}
\email{adan@us.es}
\affiliation{Departamento de F\'{\i}sica
Aplicada II, Universidad de Sevilla, 41012 Sevilla, Spain}
\date{\today}


\begin{abstract}
A proof of Bell's theorem without inequalities is presented in
which distant local setups do not need to be aligned, since the
required perfect correlations are achieved for any local rotation
of the local setups.
\end{abstract}


\maketitle


\section{Introduction}


For the last 10 years, Asher has always been right there to help
me with his advice and knowledge. He has always been reachable by
e-mail to give a quick and precise answer to all my queries with a
patience and enthusiasm far beyond that required by common
courtesy between scientists. It is therefore an honor for me to be
here as a friend to celebrate his 70th birthday. For this occasion
I have chosen to speak about a recurring subject in our
correspondence, namely, Bell's theorem, in light of one of Asher's
observations, that is to say, that the sharing of reference frames
between distant observers is an expensive resource.

Usual proofs of Bell's theorem assume that the distant observers
who perform spacelike separated measurements {\em share a common
reference frame} so that they can prepare and measure spin
components relative to it. However, a common reference frame is
not a free preexisting element in a communication scenario.
Establishing a perfect alignment between local reference frames
{\em requires the transmission of an infinite amount of
information}. Therefore, a common reference frame should be
considered an expensive resource. Indeed, the search for optimal
strategies to establish a common direction or Cartesian frame
between distant observers to any desired accuracy has attracted
much interest~\cite{PS01a,PS01b,BBBMT01,BBM01a,BBM01b,PS02}.

Addressing the question of whether or not it is possible to
develop a proof of Bell's theorem without assuming a common
reference frame seems to be relevant from a different point of
view. In 1988 Yuval Ne'eman argued that the answer to the puzzle
posed by Bell's theorem was to be found in the implicit assumption
that the detectors were aligned. Ne'eman apparently believed that
the two detectors were connected through the space-time affine
connection of general relativity~\cite{Neeman}.

However, from a practical point of view, that is, for an
experiment to show the violation of a Bell's inequality, the fact
that a {\em perfect} alignment between the source of the entangled
states and the setups of the distant measurements is difficult to
achieve is not particularly relevant, since perfect alignment is
not essential for such a test, because a violation can be obtained
even with an approximate alignment. However, in the so-called
proofs of Bell's theorem ``without
inequalities''~\cite{GHZ89,Mermin90c,Hardy92,Hardy93,Cabello01a,Cabello01b}
perfect alignment seems to be essential. These proofs are based on
inferences motivated by Einstein-Podolsky-Rosen's (EPR's)
condition for elements of reality:
\begin{quote}
{\em ``If, without in any way disturbing a system, we can predict
with certainty (i.e., with a probability equal to unity) the value
of a physical quantity, then there exist an element of physical
reality corresponding to this physical quantity''} \cite{EPR35}.
\end{quote}
These inferences arrive at a contradiction. In these
proofs, any imperfection in the required alignments leads to the
disappearance of the required inferences.

The term ``unspeakable information'' was coined by Peres and
Scudo~\cite{PS02} to designate information that cannot be
represented by a sequence of discrete symbols, such as a direction
in space or a reference frame. Here we shall show that there is a
method to prove Bell's theorem without inequalities without it
being necessary that the observers share unspeakable information
or, more precisely, without distant local setups needing to be
aligned.

The proof is based on the fact that the required perfect
correlations occur {\em for any local rotation of the local
setups}. Therefore, we shall show not only that the apparently
innocent EPR's elements of reality are incompatible with quantum
mechanics, but even that an apparently more innocent criterion for
elements of reality, namely,
\begin{quote}
{\em ``If, without in any way disturbing a system, we can predict
with certainty (i.e., with a probability equal to unity) the value
of a physical quantity}, independently of the orientation of the
measurement apparatus used, {\em then there exist an element of
physical reality corresponding to this physical quantity''},
\end{quote}
is also incompatible with quantum mechanics.


\section{Proof}


Let us consider a source emitting systems of eight qubits prepared
in the state
\begin{equation}
|\eta\rangle =
(|\phi_0\phi_0\rangle+\sqrt{3}|\phi_0\phi_1\rangle+\sqrt{3}|\phi_1\phi_0\rangle)/\sqrt{7},
\label{H1}
\end{equation}
where~$|\phi_0\rangle$ and~$|\phi_1\rangle$ are the two singlet
states obtained adding up four spin-$\frac{1}{2}$ momenta,
\begin{eqnarray}
|\phi_0\rangle & = & {1 \over 2}
(|0101\rangle-|0110\rangle-|1001\rangle+|1010\rangle), \label{P0}
\\
|\phi_1\rangle & = & {1 \over 2 \sqrt{3}}
(2|0011\rangle-|0101\rangle-|0110\rangle-|1001\rangle \nonumber \\ & &
-|1010\rangle+2|1100\rangle). \label{P1}
\end{eqnarray}
These states were introduced by Kempe {\em et al.\ }in the context
of decoherence-free fault-tolerant universal quantum
computation~\cite{KBLW01}. Let us suppose that the first four
qubits prepared in~$|\eta\rangle$ fly to Alice and the second four
qubits fly to a distant observer, Bob. On her/his four qubits,
each observer randomly chooses to measure either~$F$ or~$G$,
defined as
\begin{eqnarray}
F & = & -|\phi_0\rangle\langle\phi_0|+|\phi_1\rangle\langle\phi_1|, \\
G & = &-|\psi_0\rangle\langle\psi_0|+|\psi_1\rangle\langle\psi_1|,
\end{eqnarray}
where~$|\psi_0\rangle$ and~$|\psi_1\rangle$ are obtained,
respectively, from~$|\phi_0\rangle$ and~$|\phi_1\rangle$, by
permuting qubits~2 and~3, i.e.,
\begin{eqnarray}
|\psi_0\rangle & = & {1 \over 2} (|0011\rangle-|0110\rangle-|1001\rangle+|1100\rangle) \nonumber \\
& = & {1 \over 2} \left(|\phi_0 \rangle + \sqrt{3} |\phi_1\rangle\right), \label{S0} \\
|\psi_1\rangle & = & {1 \over 2 \sqrt{3}}
(-|0011\rangle+2|0101\rangle-|0110\rangle-|1001\rangle \nonumber \\
& & +2|1010\rangle-|1100\rangle) \nonumber \\
& = & {1 \over 2} \left( \sqrt{3} |\phi_0 \rangle - |\phi_1 \rangle\right). \label{S1}
\end{eqnarray}
The observable~$F$ ($G$) has three possible outcomes:~$-1$,
corresponding to~$|\phi_0\rangle$ ($|\psi_0\rangle$), $1$
corresponding to~$|\phi_1\rangle$ ($|\psi_1\rangle$), and~$0$,
which never occurs because the local subsystems have total spin
zero. Measuring~$F$ is thus equivalent to distinguishing with
certainty between~$|\phi_0\rangle$ and~$|\phi_1\rangle$ with a
single test on the four qubits, and measuring~$G$ is equivalent to
distinguishing with certainty between~$|\psi_0\rangle$
and~$|\psi_1\rangle$. Alice's measurements on qubits~1 to~4 are
assumed to be spacelike separated from Bob's measurements on
qubits~5 to~8.

The state~$|\eta\rangle$ can also be expressed as
\begin{eqnarray}
|\eta\rangle & = & (4
|\phi_0\psi_0\rangle+\sqrt{3}|\phi_1\psi_0\rangle+3|\phi_1\psi_1\rangle)/2\sqrt{7}
\label{H2} \\ & = & (4
|\psi_0\phi_0\rangle+\sqrt{3}|\psi_0\phi_1\rangle+3|\psi_1\phi_1\rangle)/2\sqrt{7}
\label{H3} \\ & = & (7
|\psi_0\psi_0\rangle+3\sqrt{3}|\psi_0\psi_1\rangle+3\sqrt{3}|\psi_1\psi_0\rangle
\nonumber \\ & & -3|\psi_1\psi_1\rangle)/4\sqrt{7}. \label{H4}
\end{eqnarray}
Moreover, since~$|\phi_0\rangle$, $|\phi_1\rangle$,
$|\psi_0\rangle$, and~$|\psi_1\rangle$ are invariant under the
tensor product of four equal unitary operators, then they are
invariant under local rotations. Therefore,
expressions~(\ref{H1}) and (\ref{H2})--(\ref{H4})
remain unchanged after local rotations. Consequently, if~$R_A$
and~${\cal R}_A$ ($R_B$ and~${\cal R}_B$) are rotations of Alice's
(Bob's) setups for measuring, respectively, $F$ and~$G$ relative
to the reference frame of the source then, in the
state~$|\eta\rangle$, for {\em any} rotations~$R_A$, ${\cal R}_A$,
$R_B$, and~${\cal R}_B$,
\begin{eqnarray}
P(R_A F =1,R_B F=1) & = & 0, \label{G4} \\
P(R_A F =1\,|\,{\cal R}_B G=1) & = & 1, \label{G3} \\
P(R_B F =1\,|\,{\cal R}_A G=1) & = & 1, \label{G2} \\
P({\cal R}_A G=1,{\cal R}_B G=1) & = & {9 \over 112}, \label{G1}
\end{eqnarray}
where~$P(R_A F =1,R_B F=1)$ is the joint probability that both
Alice and Bob obtain the outcome~$1$ when both perform
experiment~$F$ (or any experiment consisting on independently
rotating their setups for measuring~$F$), and~$P(R_A F
=1\,|\,{\cal R}_B G=1)$ is the probability that Alice obtains the
outcome~$1$ when she performs experiment~$F$ (or any experiment
consisting on rotating her setup for measuring~$F$), conditioned
to Bob obtaining the outcome~$1$ when he performs experiment~$G$
(or any experiment consisting on rotating his setup for
measuring~$G$).

From property (\ref{G1}), if both Alice and Bob choose the setup
for measuring~$G$, then in~$8\%$ of the events the outcome is~$1$
in both cases. This is true even if Alice applies any
rotation~${\cal R}_A$ to her setup and Bob applies any
rotation~${\cal R}_B$ to his setup.

From property (\ref{G2}), if Alice measures~$G$ and obtains the
outcome~$1$, then she can predict with certainty that, if Bob
measures~$F$, he will obtain~$1$. According to Einstein, Podolsky,
and Rosen (EPR), this fact must be interpreted as sufficient
evidence that there is a local ``element of reality'' in Bob's
qubits determining this outcome~\cite{EPR35}. Moreover, EPR
reasoning seems to be even more inescapable in our example, since
Alice's prediction with certainty is valid even if Alice applies
any rotation~${\cal R}_A$ to her setup for measuring~$G$ and Bob
applies any rotation~$R_B$ to his setup for measuring~$F$.

Analogously, from property (\ref{G3}), if Bob measures~$G$
(or~${\cal R}_A G$) and obtains~$1$, then he can predict with
certainty that, if Alice measures~$F$ (or~$R_A F$), she will
obtain~$1$. Again, according to EPR, there must be a local element
of reality in Alice's qubits determining this outcome.

\begin{figure}
\centerline{\includegraphics[width=8.2cm]{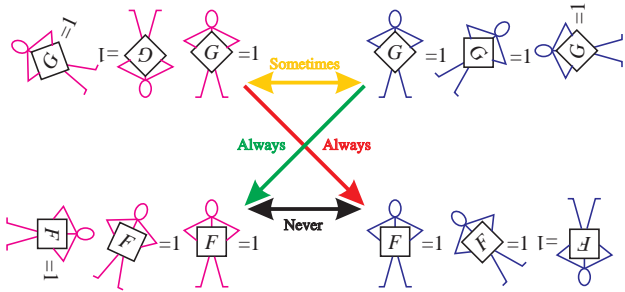}}
\caption{\label{dib13} Let us suppose that Alice (up in the
left-hand side) measures $G$ and obtains the result $1$. Then, if
Bob (down in the right-hand side), who is spacelike separated from
Alice, measures $F$, he {\em always} obtains $1$, even if she/he
rotates her/his measurement apparatus. Analogously, if Bob (up in
the right-hand side) measures $G$ and obtains $1$, then he can
predict that, if Alice (down in the left-hand side) measures $F$,
she {\em always} obtains $1$, even if Alice/Bob rotates her/his
apparatus. If Alice and Bob measure $G$, {\em sometimes} (in $8\%$
of the cases) they both obtain $1$. In those cases, what if,
instead of measuring $G$, they had measured $F$?. If EPR's
elements of reality do exist, then, at least in $8\%$ of the
cases, both of them would have obtained $F=1$. However, they {\em
never} both obtain 1.}
\end{figure}

Therefore, assuming EPR's point of view, for at least~$8\%$ of the
systems prepared in the state~$|\eta\rangle$, there must be two
joint local elements of reality: one for Alice's qubits,
corresponding to~$R_A F=1$, and one for Bob's qubits,
corresponding to~$R_B F=1$. However, this inference is in
contradiction with property (\ref{G4}), which states that the
joint probability of obtaining the outcomes~$R_A F = 1$ and~$R_B
F= 1$ is zero. The logical structure of the proof is summarized in
Fig.~\ref{dib13}.

This is a simple and powerful proof that the concept of element of
reality, as defined by EPR, is incompatible with quantum
mechanics, even if the predictions with certainty are valid not
only for a particular alignment of the distant setups but for {\em
any} possible rotation.


\section{Remarks}


The fact that the required perfect correlations occur for any local
rotation of the local setups arises from the fact that
the state $|\eta\rangle$ is invariant under local rotations,
\begin{equation}
R_A \otimes R_B |\eta\rangle = |\eta\rangle,
\end{equation}
This
property comes from the fact that the states $|\phi_0\rangle$ and
$|\phi_1\rangle$ are invariant under the same unitary
transformation applied to the four qubits,
\begin{equation}
U \otimes U \otimes U \otimes U |\phi_j\rangle =
|\phi_j\rangle,
\end{equation}
where $U$ is any unitary operation on one qubit. This invariance
of the four-qubit states $|\phi_0\rangle$ and $|\phi_1\rangle$ has
been recently demonstrated in a laboratory using four-photon
polarization-entangled states produced by spontaneous parametric
down-conversion~\cite{BEGKCW03}.

A practical advantage and a remarkable property of this proof is
that measuring~$F$ or~$G$ does not require collective measurements
on two or more qubits but instead a single test on each of the
four qubits. Measuring~$F$ is equivalent to distinguishing
between~$|\phi_0\rangle$ and~$|\phi_1\rangle$ with a single test.
Remarkably, the only two orthogonal states invariant under any
tensor product of four equal unitary operators that can be
reliably distinguished by fixed measurements on the four
individual qubits are~$|\phi_0\rangle$ and~$|\phi_1\rangle$ and
those obtained from them by permuting qubits (such
as~$|\psi_0\rangle$ and~$|\psi_1\rangle$)~\cite{Cabello03a}.

To distinguish with certainty between~$|\phi_0 \rangle$
and~$|\phi_1 \rangle$, it is enough to measure the spin component
of the first two qubits along the same direction and the spin
component of the other two qubits along a perpendicular direction.
This can be seen by resorting to the invariance under any tensor
product of four equal unitary operators and expressing these
states in the basis of eigenstates of~$\sigma_{z1} \otimes
\sigma_{z2} \otimes \sigma_{x3} \otimes \sigma_{x4}$,
\begin{eqnarray}
|\phi_0 \rangle & = & {1 \over 2}
(-|01\bar{0}\bar{1}\rangle+|01\bar{1}\bar{0}\rangle+|10\bar{0}\bar{1}\rangle-|10\bar{1}\bar{0}\rangle),
\label{lp0}
\\
|\phi_1 \rangle & = & {1 \over 2 \sqrt{3}}
(|00\bar{0}\bar{0}\rangle-|00\bar{0}\bar{1}\rangle-|00\bar{1}\bar{0}\rangle+|00\bar{1}\bar{1}\rangle \nonumber \\
& & -|01\bar{0}\bar{0}\rangle+|01\bar{1}\bar{1}\rangle
-|10\bar{0}\bar{0}\rangle+|10\bar{1}\bar{1}\rangle \nonumber \\ &
&
+|11\bar{0}\bar{0}\rangle+|11\bar{0}\bar{1}\rangle+|11\bar{1}\bar{0}\rangle+|11\bar{1}\bar{1}\rangle),
\label{lp1}
\end{eqnarray}
where~$\sigma_z |0\rangle=|0\rangle$, $\sigma_z
|1\rangle=-|1\rangle$, $\sigma_x |\bar{0}\rangle=|\bar{0}\rangle$,
$\sigma_x |\bar{1}\rangle=-|\bar{1}\rangle$ [$|\bar{0}\rangle
=(|0\rangle+|1\rangle)/\sqrt{2}$ and~$|\bar{1}\rangle
=(|0\rangle-|1\rangle)/\sqrt{2}$]. According to~(\ref{lp0})
and~(\ref{lp1}), if the measurements on the individual qubits
are~$\sigma_{z1}$, $\sigma_{z2}$, $\sigma_{x3}$, $\sigma_{x4}$ (or
any rotation thereof), then, among the $16$ possible outcomes,
four occur (with equal probability) only in the state~$|\phi_0
\rangle$, and the other twelve occur (with equal probability) only
in the state~$|\phi_1 \rangle$ (this has been experimentally
demonstrated in~\cite{BEGKCW03}). Therefore, to measure~$F$ ($G$),
it is enough to measure the spin component of qubits~$1$ and~$2$
($1$ and~$3$) along the same direction and the spin component of
the other two qubits along a perpendicular direction.

On the other hand, fixed measurements on the four individual
qubits are not enough for a Greenberger-Horne-Zeilinger-like proof
without unspeakable information. Such a proof requires conditioned
measurements on the four individual qubits~\cite{Cabello03b}.


\section{Conclusions}


What conclusions should we draw from all this? David Mermin would say:
\begin{quote}
``Some people conclude from all this that the character of {\em
your} [or Alice's] stuff can indeed be altered by decisions made
by your far away friend [Bob]. This is called quantum nonlocality.
Others, myself among them, conclude that it is treacherous to make
judgments about the character of your stuff, and extremely
treacherous to reason from what actually happened to what might
have happened but didn't'' \cite{Mermin03}.
\end{quote}

Which is a less concise way of saying what Asher perfectly summarized 26
years ago:
\begin{quote}
``Unperformed experiments have no results''
\cite{Peres78}.
\end{quote}


\section{Acknowledgements}


The author thanks the organizers of the conference {\em Recent
Developments in Quantum Physics}, specially Ady Mann and Liz
Youdim, and the Spanish Ministerio de Ciencia y Tecnolog\'{\i}a
Project~BFM2002-02815 for support.



\begin{thebibliography}{99}


\bibitem{PS01a}
A. Peres and P.F. Scudo,
Phys. Rev. Lett. {\bf 86}, 4160 (2001).

\bibitem{PS01b}
A. Peres and P.F. Scudo,
Phys. Rev. Lett. {\bf 87}, 167901 (2001).

\bibitem{BBBMT01}
E.~Bagan, M.~Baig, A.~Brey, R.~Mu\~{n}oz-Tapia, and R.~Tarrach,
Phys. Rev. A {\bf 63}, 052309 (2001).

\bibitem{BBM01a}
E. Bagan, M. Baig, and R. Mu\~{n}oz-Tapia,
Phys. Rev. A {\bf 64}, 022305 (2001).

\bibitem{BBM01b}
E. Bagan, M. Baig, and R. Mu\~{n}oz-Tapia,
Phys. Rev. Lett. {\bf 87}, 257903 (2001).

\bibitem{PS02}
A. Peres and P.F. Scudo,
in {\em Quantum Theory: Reconsideration of Foundations}, edited by
A. Khrennikov (V\"{a}xj\"{o} University Press, V\"{a}xj\"{o},
Sweden, 2002).


\bibitem{Neeman}
N.D. Mermin (private communication).


\bibitem{EPR35}
A. Einstein, B. Podolsky, and N. Rosen,
Phys. Rev. {\bf 47}, 777 (1935).


\bibitem{GHZ89}
D.M. Greenberger, M.A. Horne, and A. Zeilinger,
in {\em Bell's Theorem, Quantum Theory, and Conceptions of the
Universe}, edited by M. Kafatos (Kluwer, Dordrecht, 1989), p. 69.

\bibitem{Mermin90c}
N.D. Mermin,
Phys. Rev. Lett. {\bf 65}, 3373 (1990).

\bibitem{Hardy92}
L. Hardy,
Phys. Rev. Lett. {\bf 68}, 2981 (1992).

\bibitem{Hardy93}
L. Hardy,
Phys. Rev. Lett. {\bf 71}, 1665 (1993).

\bibitem{Cabello01a}
A. Cabello,
Phys. Rev. Lett. {\bf 86}, 1911 (2001).

\bibitem{Cabello01b}
A. Cabello,
Phys. Rev. Lett. {\bf 87}, 010403 (2001).


\bibitem{KBLW01}
J. Kempe, D. Bacon, D.A. Lidar, and K.B. Whaley,
Phys. Rev. A {\bf 63}, 042307 (2001).


\bibitem{BEGKCW03}
M.~Bourennane, M.~Eibl, S.~Gaertner, C.~Kurtsiefer, A.~Cabello,
and H.~Weinfurter, e-print quant-ph/0309041 (to be published in
Phys. Rev. Lett.).


\bibitem{Cabello03a}
A. Cabello,
Phys. Rev. Lett. {\bf 91}, 230403 (2003).

\bibitem{Cabello03b}
A. Cabello,
Phys. Rev. A {\bf 68}, 042104 (2003).

\bibitem{Mermin03}
N.D. Mermin,
Am. J. Phys. {\bf 71}, 296 (2003).

\bibitem{Peres78}
A. Peres,
Am. J. Phys. {\bf 46}, 745 (1978).


\end{thebibliography}
\end{document}